\documentclass[aps,prl,reprint,groupedaddress]{revtex4-2}
\usepackage{graphicx}
\usepackage{txfonts}
\usepackage[geometry]{ifsym}

\begin{document}


\title{Crystal structures and  superconducting properties of metallic double-chain based cuprate   Pr$_{2}$Ba$_{4}$Cu$_{7}$O$_{15-\delta}$ }

\author{Masahide Hagawa$^1$,  Michiaki Matsukawa$^1$\thanks{matsukawa@iwate-u.ac.jp}, Kota Niinuma$^1$, Reiya Kudo$^1$,  Yuto Mizusihima$^1$,  Naohisa Kawarada$^1$,  Hajime Yamamoto$^2$,   Kazuhiro Sano$^3$,  Yoshiaki Ōno$^4$ and Takahiko Sasaki$^5$}


\affiliation{$^1$Faculty of Science and Engineering, Iwate University, Morioka 020-8551, Japan \\
$^2$Institute of Multidisciplinary Research for Advanced Materials, Tohoku University, Sendai 980-8577, Japan\\
$^3$Department of Physics Engineering, Mie University, Tsu 514-8507, Japan \\
$^4$Department of Physics, Niigata University, Niigata 950-2181, Japan \\
$^5$Institute for Materials Research, Tohoku University, Sendai 980-8577, Japan}


\date{\today}

\begin{abstract}
We demonstrated the lattice structures and  the superconducting phases of metallic double-chain based cuprate   Pr$_{2}$Ba$_{4}$Cu$_{7}$O$_{15-\delta}$ exhibiting higher $T\mathrm{_{c}}$.  
 After the oxygen heat treatment on  citrate pyrolysis precursors, their reduction treatment followed by a quench procedure caused higher $T\mathrm{_{c}}$ samples with 26.5$\sim$30 K.
The crystal structural parameters for 
the superconducting sample   ($\delta$ = 0.81) were analyzed from the powder synchrotron X-ray diffraction data using  RIETAN-FP program.
The  effect of magnetic field on the superconducting phase  of these samples with different oxygen defects  ($\delta$ =0.73, 0.81 and 0.87) was examined, for our understanding of  the superconducting magnetic field-temperature  phase diagram.  
For the $\delta$ = 0.87 sample  with $T\mathrm{_{c}}$ $\sim$30 K,  the resistive critical field  $H\mathrm{_{c}}^{*}$ was estimated to be 13 T at 4.2 K.   
 The oxygen deficiency  dependence  on $T\mathrm{_{c, on}}$ for our samples   was  compared with the data of several other groups.
\end{abstract}


\maketitle

\section{Introduction}
\begin{figure}[ht]
\begin{center}
\includegraphics[width=7cm]{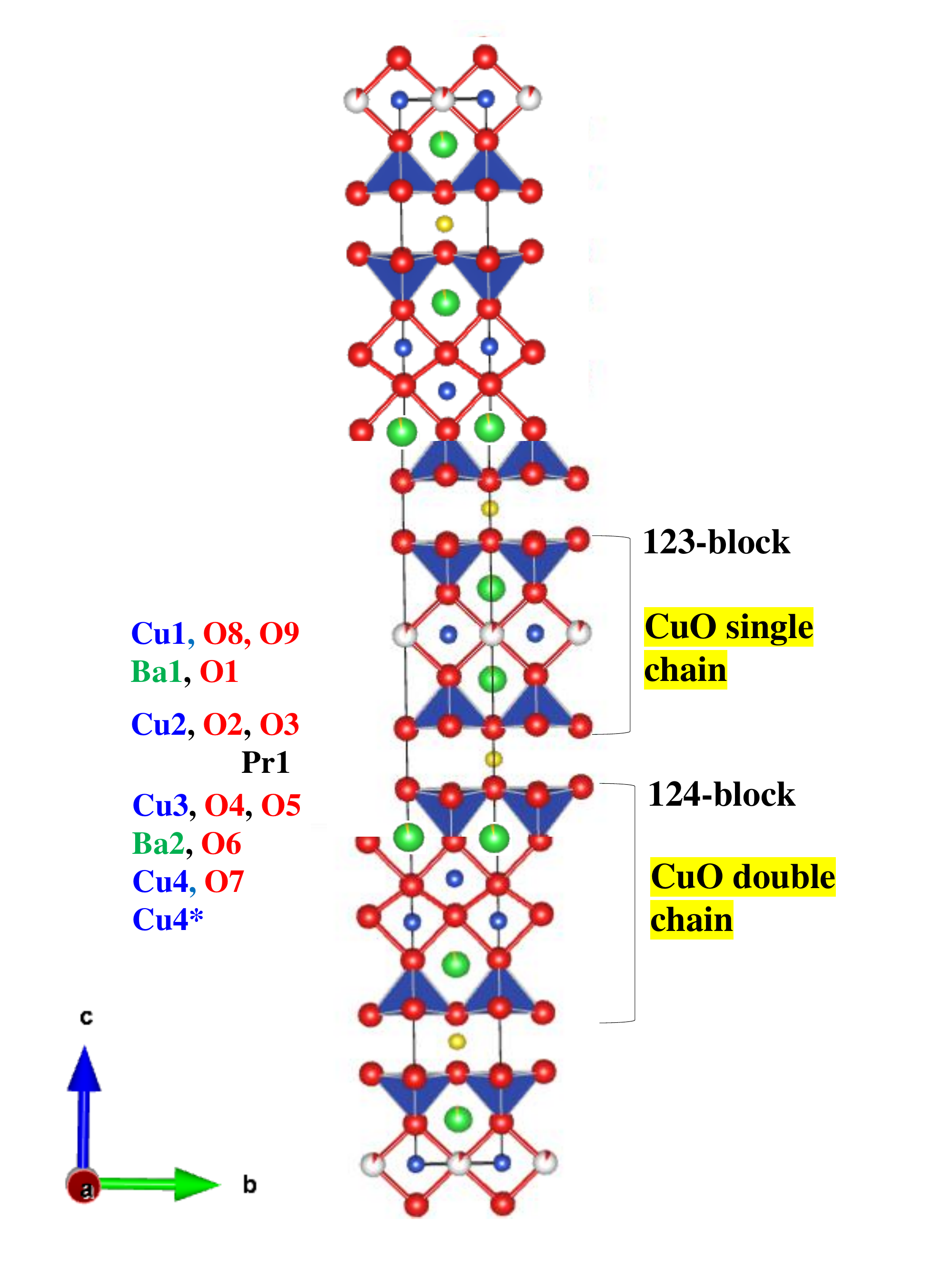}
\caption{Projection along (100) of  the crystal structure of Pr$_{2}$Ba$_{4}$Cu$_{7}$O$_{15-\delta}\ $. The CuO sinlgle-chain 123- and double-chain 124-blocks are alternatively stacked along the $c$-axis. 
The nomenclature of atomic positions  is displayed on the left hand side. The polyhedral planar part represents CuO$_{2}$ plane site.}  
\label{CR}
\end{center}
\end{figure} 

\begin{figure*}[ht]
\begin{center}
\includegraphics[width=12cm]{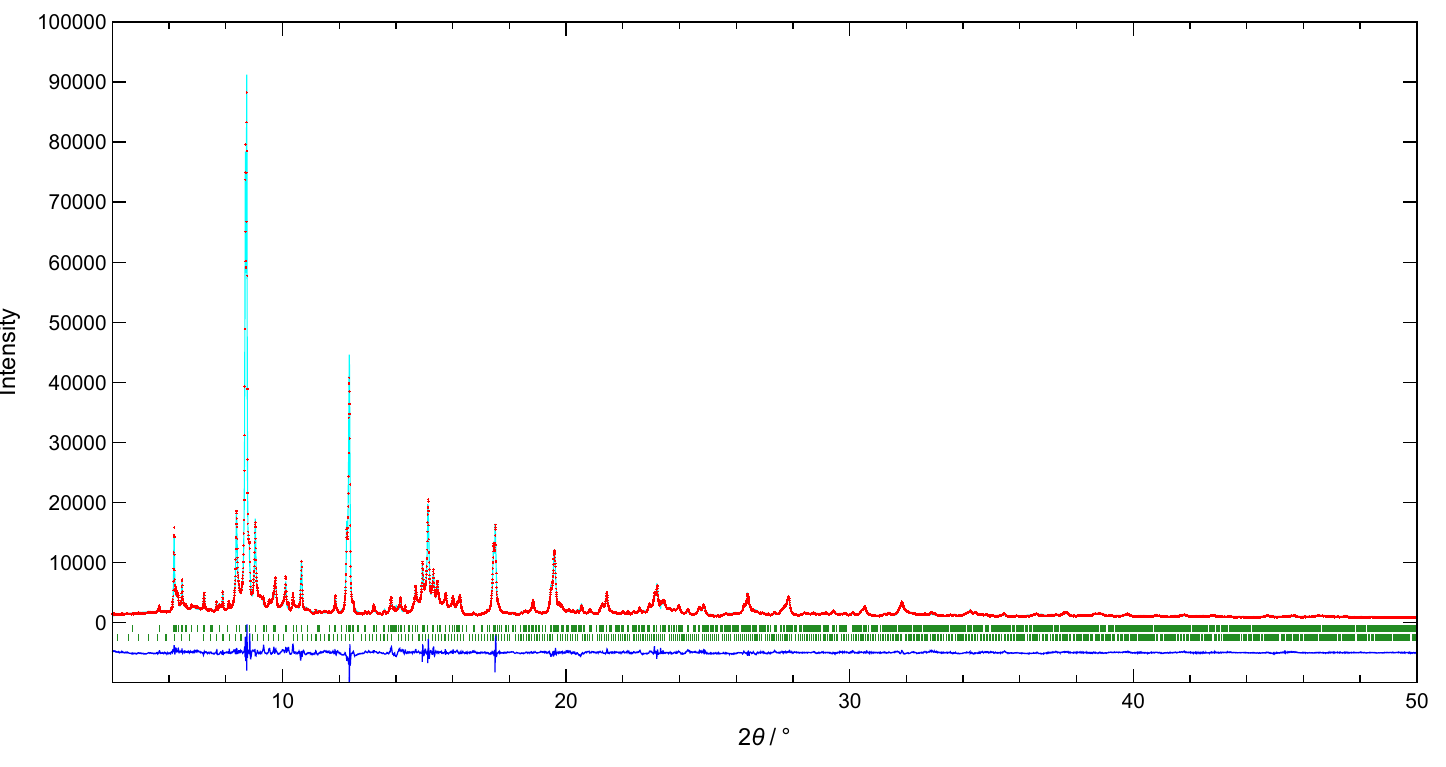}
\caption{Powder synchrotron X-ray diffraction pattern on the superconducting sample Pr$_{2}$Ba$_{4}$Cu$_{7}$O$_{15-\delta}  $  ($\delta$ = 0.81) at room temperature.  The calculated and residual curves are also shown using  RIETAN-FP program  for  two-phase Rietveld refinement.  The upper and lower tick marks represent peak positions for the Pr$_{2}$Ba$_{4}$Cu$_{7}$O$_{15-\delta}  $ and BaCuO$_{2}$ phases, respectively. The mass fractions of Pr247 and BaCuO$_{2}$  were estimated to be $ 96.0 \% : 4.0 \% $. $\lambda$ = 0.42 \AA.  }
\label{RI} 
\end{center}
\end{figure*}

Since the discovery of superconductivity in the metallic double-chain based cuprate   Pr$_{2}$Ba$_{4}$Cu$_{7}$O$_{15-\delta}$ (Pr247)\cite{MA04}, experimental and  theoretical studies on this system have been  extensively  reported \cite{YA05,HA06,WA08,TA13,SA05,NA07,OK07}.  The Pr247  system is considered to be an intermediate compound between the 2D high- $T\mathrm{_{c}}$ layered cuprate superconductors and the quasi-1D organic superconductors, as well as the spin-ladder compounds.
It is well known that the spin ladder system exhibits  superconductivity with $T\mathrm{_{c}}\sim$12 K only  under an external pressure above $\sim$3.0 GPa \cite{UE96}.   Pr$_{2}$Ba$_{4}$Cu$_{7}$O$_{15-\delta}$ cuprate is an electron-doped superconductor with  $T\mathrm{_{c}}\sim$15 K at ambient pressure through  controlling the amount of oxygen deficiency due to a reduction treatment. 

For the crystal structure of Pr$_{2}$Ba$_{4}$Cu$_{7}$O$_{15-\delta}\ $ (Fig.\ref{CR}),  the PrBa$_{2}$Cu$_{3}$O$_{7-\delta}$ (Pr123)-type blocks containing the CuO sinlgle chains are alternatively stacked along the $c$-axis with   the  PrBa$_{2}$Cu$_{4}$O$_{8}$ (Pr123)-type ones containing the CuO  double chains. 
The Pr123 and Pr124 compounds show no superconductivity associated with the CuO$_{2}$ planes  because the strong hybridization between Pr-4f and O-2p orbitals suppresses their metallic conduction \cite{FE93}.  Furthermore, an angle-resolved photoemission study revealed  the CuO single chain in Pr123  and the CuO double chain in Pr124 exhibit  insulating and metallic characters, respectively, although both chains are approximately quarter filling \cite{MI00}. 
 Recently,  a Cu nuclear quadrupole resonance study on Pr247 with a full volume fraction of superconductive phase has revealed that the CuO$_{2}$ plane is an antiferromagnetically insulating state at 2 K despite its bulk superconductivity\cite{NI22,WA05}.
 This finding strongly indicates that the superconductivity in Pr247 appears along the metallic CuO double chains, considering the oxygen deficiency at the CuO single chain site. 
 For our further understanding of  dimensionality in the electronic state of CuO double chains, the lattice, transport, and magnetic properties of Pr247 have  been examined under an external pressure. 
Both the resistivity and magnetization measurements revealed that  the superconducting behaviors in Pr247 are suppressed by the pressure effect\cite{IS09,KU16}.  The negative pressure dependences of the superconducting phase  are qualitatively explained by the normal to superconductive phase diagram of CuO double chains on the basis of Tomonaga-Luttinger liquid theory\cite{SA05}. 
The magneto resistance (MR) effect of the normal phase of Pr247  is enhanced due to applied pressure, which is probably  related to a  warped Fermi surface  induced by the pressure. 
 Taniguchi et al, found out from in-situ XRD experiments under pressure that the $a$-axis lattice parameter  across the CuO double chain along  the  $b$-axis is elongated against pressure above 2.0 GPa\cite{TA21}.

\begin{figure*}[ht]
\begin{center}
\includegraphics[width=14cm]{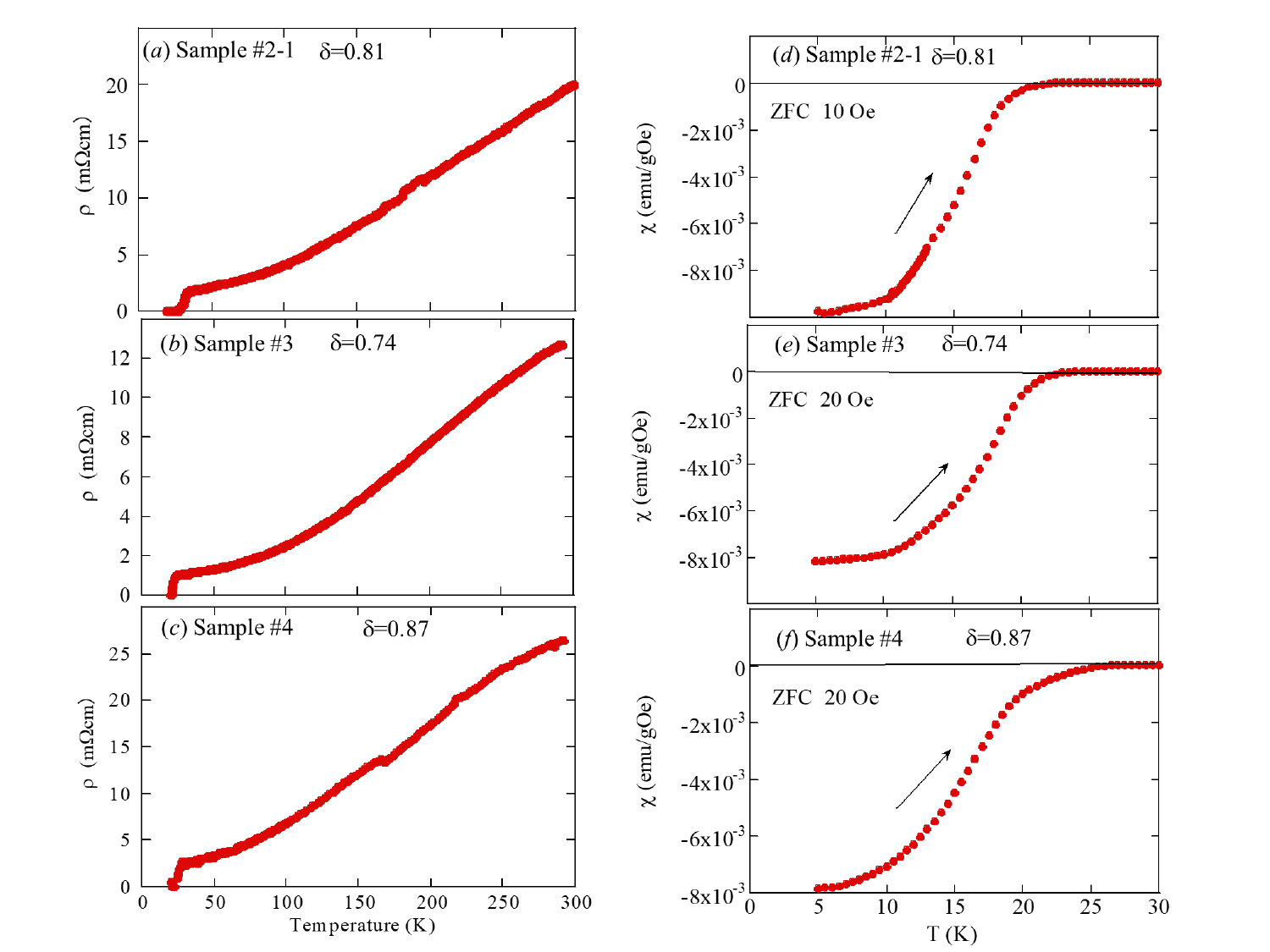}
\caption{Temperature dependences of  electric resistivities  for the superconducting Pr$_{2}$Ba$_{4}$Cu$_{7}$O$_{15-\delta }$. ($a$) sample \#2-1 with  $\delta$=0.81, ($b$) sample \#3 with $\delta$=0.74, and ($c$) sample \#4 with $\delta$=0.87.  In the inset,  the low-temperature magnetization curves measured under a zero field cooling (ZFC) scan  are  plotted, to clarify a bulk superconductivity.   }
\label{RT}
\end{center}
\end{figure*}

\begin{table*}[htb]
\caption{Lattice parameters and superconducting properties  for  the as-sintered and oxygen-reduction samples of   Pr$_{2}$Ba$_{4}$Cu$_{7}$O$_{15-\delta}  $  ($\delta$ =0, 0.73, 0.81, and 0.87). 
In details, see the corresponding text and references.   $f_\mathrm{SC}$: the superconducting volume fraction estimated from the ZFC magnetization data at low temperatures. $\mathrm{T_{c, on}}$: defined as the onset temperature of resistive drop. The reliability factors for four samples are listed. 
 For the oxygen defect  samples with $\delta$ =0.73, 0.81, and 0.87,  the dimension of  their rectangular shape  and their relative density ($d$) are also given.  }
\begin{center}
\begin{tabular}{lllllllllll}
\hline \hline
{Sample} & {oxygen} &\ \ ${a}$& \ \ ${b}$ & \ \ ${c}$ &  $f\mathrm{_{SC}}$ &  $\mathrm{T_{c, on}}$ &   \ \ $R\mathrm{_{wp}}$&$R\mathrm{_{p}}$&$R\mathrm{_{B}}$    \\
 No.  &  {defect $\delta$}  & \ \ (\AA) & \ \ (\AA) & \ \ (\AA)  & $$ (\%)&($K$)  &(\%)&(\%)&(\%) \\ 

\hline
\#1 & {-}& { 3.87958 (5)}  &{3.89975 (1)}& 50.6281(2)&\ \ {-}  &  \ \ {-}   &7.95&5.61& 1.54\\
\#2-0 & {-} &{3.8791(5)}*& {3.8997(5)} *& 50.628(6)*&\ \ {-}  &  \ \ {-}   & 11.9*&  {9.14*}&{4.24*}\\

\#2-1 & {0.81} & 3.88287(4)& {3.89003(4)} &50.8485(4) &{63}&{28.0
}  &6.62&4.86&2.83\\ 
&$1.1\times3.7$mm$^{2}$ & $4.2$ mm&$d=89\%$  \\
\#3 & {0.73} & { 3.8821(1)} &3.8899(1)&50.8160(6) &{59}&{26.5
}   &4.88&3.69&4.36\\
&$0.8\times4.1$mm$^{2}$ &5.7  mm&$d=96\%$  \\
\#4 & {0.87} & {-} & -&- &{53}&{30.0
}   &-&-&-\\ 
&$2.1\times5.0$ mm$^{2}$ & 5.5 mm&$d=84\%$  \\


\hline 
ref. \cite{YA05} & {-} &{3.89127(9)}& {3.90714(8)} & 50.695(1)&\ \ {-}  &  \ \ {-}   &4.51&3.31&  \ \ {-} \\
\hline \hline
*Ultima & IV&  Cu-K$\alpha $  \\
\label{TA1}
\end{tabular}
\end{center}
\end{table*}

In this paper,  we demonstrate crystal structures and  superconducting properties of metallic double-chain based cuprate   Pr$_{2}$Ba$_{4}$Cu$_{7}$O$_{15-\delta}$.  
In the next section,  the experimental outline is described.  In Section 3, we present the powder synchrotron X-ray diffraction data  on the superconducting sample Pr$_{2}$Ba$_{4}$Cu$_{7}$O$_{15-\delta}  $  ($\delta$ = 0.81)  using  RIETAN-FP program  for  two-phase  Pr$_{2}$Ba$_{4}$Cu$_{7}$O$_{15-\delta} $ and BaCuO$_{2} $.  We examine the magnetic field effect on the superconducting phase  of these three samples with different oxygen defects  ($\delta$ =0.73, 0.81 and 0.87).  The superconducting magnetic field-temperature  phase diagram is  determined from the low-temperature MR effect.  
 Concluding remarks are given in the final section.

\section{Experiment}
Precursors for the  superconducting cuprates Pr$_{2}$Ba$_{4}$Cu$_{7}$O$_{15-\delta}$  were prepared through spontaneous combustion reactions using a  citrate pyrolysis method\cite{HA06,SE22}. (refer to a video file recording the typical self-ignition process attached in Appendix A in ref.\cite{TE22}). 
The fine powders crushed from their precursors were pressed  to form  a disk pellet and its pellet was cut into several  pieces.   These smaller pieces from  the original pellet  were initially  set  in a cage of  Ag fine meshes and were then 
annealed under ambient oxygen atmosphere at 890-891 $^{\circ} $C for 96-110 hours in an electric tube furnace.  A narrow stability range of annealing temperature between 890 and 891 $^{\circ} $C allowed synthesis of pure phase of Pr247 with the aid of a three-zone temperature controller.  
 As for the as-sintered  samples,    oxygen removed  procedures are further needed in order to realize their superconducting phases.  The oxygen contents were reduced through heat treatment  in a vacuum atmosphere at 500-600  $^{\circ} $C for  3-5 days.  We conducted gravimetric analysis on  the present three samples  and their  oxygen defects  were estimated to be $\delta=$ 0.73, 0.81, and 0.87 (referred as  samples \#2-1, \#3, and \#4). 
 An electron probe micro  analysis on the as-sintered samples prepared under a similar oxygen annealing condition   showed the average  oxygen content = $\sim 14.96$. 
A quench process in air from 300  $^{\circ} $C down to room temperature was an effective approach to achieve better superconducting properties, in comparison to furnace cooling procedures in our previous synthesis.  
For the $\delta$ =0.73, 0.81, and 0.87 oxygen-reduction samples,  the dimension of  their rectangular shape and their relative density ($d$) are given in Table \ref{TA1}.
We carried out powder synchrotron X-ray diffraction measurements on the fine powders crashed from the  superconducting bulk samples   at SPring-8 beamline BL02B2 at room temperature.  The wavelength of the incident X-ray, $\lambda$ = 0.42 \AA,   was utilized. The collected diffraction patterns were fitted using RIETAN-FP program   two-phase Rietveld refinement\cite{IZ07}.   
The electric resistivity measurement  was measured by the $dc$ four-terminal method. The magneto-transport up to 9 T was measured by the $ac$ four-probe method using a physical property measuring system (PPMS, Quantum Design),  from 4 K  to 40 K at several magnetic fields  ($\mu_{0}H$=0, 1, 3, 6, and 9 T) after  the zero-field-cooling (ZFC) scan . The low-temperature magneto-resistance (MR) effect   (up to 14 T) was measured in a superconducting magnet (15T-SM)  at Institute for Materials Research, Tohoku University.  The electric current  $I$ was applied longitudinally to the sample ; consequently, 
the applied magnetic field $H$ was transverse to the sample (because $H\perp I$). 
The $dc$ magnetization was performed under ZFC in a commercial superconducting quantum interference  device magnetometer (Quantum Design, MPMS).

\begin{table*}[htb]
 \caption{Atomic position parameters of  Pr$_{2.2}$Ba$_{3.8}$Cu$_{7}$O$_{15-\delta }$ ($\delta$=0.8). In our Rietveld refinement of the synchrotron  and conventional X-ray diffraction patterns for all the  samples investigated,  similar atomic position parameters are adopted. $B$  is an isotropic thermal parameter.  For oxygen atoms, $B$ is fixed to be 1.0 $\AA^{2}$.  }
  \begin{tabular}{lccccccccccll}
   \hline \hline
 \#2-1 && $\delta$=0.81&&($$Ammm$$) &&  &&\ \ \ \   &&\ \ \ \  &&   \\
Atom && Site&& Occ. &&$x$&&\ \ $y$\ \   &&\ \ $z$\ \   && $B$  ($\AA^{2}$)  \\ 
 \hline
 Pr1 &&4$j$&&1.0 && 0.5 &&0.5  &&0.11559(4) &&0.15(2)  \\ 
 Ba1 &&4$j$&&0.95&& 0.5 &&0.04188(3)  &&0.5 &&0.51(2) \\ 
  Pr2 &&4$j$&&0.05&& 0.5 &&0.04188(3)  &&0.5 &&0.15(2) \\ 
Ba2 &&4$j$&&0.95&& 0.5 &&0.18865(3) &&0.5 &&0.51(2) \\ 
Pr3 &&4$j$&&0.05&& 0.5 &&0.18865(3) &&0.5 &&0.15(2)  \\ 
Cu1&&2$a$&&1.0&& 0.0 &&0.0  &&0.0 &&0.33(2)  \\ 
Cu2&&4$i$&&1.0&& 0.0 &&0.0 && 0.08073(7)  &&0.33(2)  \\ 
Cu3&&4$i$&&1.0&& 0.0 &&0.5  && 0.15027(7)&&0.33(2)  \\ 
Cu4&&4$i$&&1.0&& 0.0 &&0.0  &&0.22965(6) &&0.33(2)  \\ 
O1&&4$i$&&1.0&& 0.0&&0.0  &&0.0386(4)&&1.0 \\ 
O2&&4$j$&&1.0&& 0.5&&0.0 &&0.0828(4)&&1.0 \\
O3&&4$i$&&1.0&& 0.0&&0.5  &&0.0861(3)&&1.0 \\ 
O4&&4$j$&&1.0&& 0.5&&0.0 &&0.1486(4)&&1.0 \\
O5&&4$i$&&1.0&& 0.0&&0.5  &&0.1421(3)&&1.0 \\ 
O6&&4$i$&&1.0&& 0.0&&0.0  &&0.1935(3)&&1.0 \\ 
O7&&4$i$&&1.0&& 0.0&&0.5  &&0.2355(3) &&1.0 \\ 
O8&&2$b$&&0.1&& 0.0&&0.5  &&0.0&&1.0 \\
O9&&2$d$&&0.1&& 0.5&&0.0  &&0.0&&1.0 \\ 
 
 \hline

 \hline \hline
  \end{tabular}
  \label{TA2}
\end{table*}

\begin{table*}[htb]

\caption{Selected bond lengths for Pr$_{2}$Ba$_{4}$Cu$_{7}$O$_{15-\delta }$. 
determined by the powder  synchrotron X-ray diffraction patterns.   See Fig. \ref{CR} for positional notation of Pr247.  The atomic positions for Pr 124 and Pr123 are referred to the published data\cite{YA01, LO90}.
Cu$_{\mathrm{pl}}$ and  Cu$_{\mathrm{ch}}$  denote the planar and chain copper sites, respectively. O$_{\mathrm{ap}}$ and  O$_{\mathrm{ch}}$ are  the apical and  double-chain oxygen sites.   }

\begin{tabular}{l|ll|ll|ll} 
\hline
bond length  &Pr247/$\delta$=0.81 &\#2-1  &Pr247 ref. \cite{YA05}&as-sintered&Pr124 &Pr123   \\
 (\AA) & 124-block& 123-block  & 124-block  &123-block&ref. \cite{YA01}&ref. \cite{LO90}    \\%

 \hline
Cu$_{\mathrm{pl}}$-Cu$_{\mathrm{pl}}$ &\ \ \ \ \ \ \ \ \ \ \ \ Cu2-Cu3 &   & \ \ \ \ \ \ \ \ \ \ \ \ Cu2-Cu3   &&Cu2-Cu2&Cu2-Cu2\\
 & \ \  \ \ \ \ \ \ \ \ \ \ 3.528(5) & &\ \ \ \ \ \ \ \ \ \ \ \ 3.525& &3.508&3.554 \\
Cu$_{\mathrm{pl}}$-O$_{\mathrm{ap}}$& Cu3-O6 &Cu2-O1    &Cu3-O6   &Cu2-O1&Cu2-O1&Cu2-O2\\
 & 2.183(4)&2.095(20)&2.196&2.136&2.234  & 2.232 \\ 
Cu$_{\mathrm{ch}}$-O$_{\mathrm{ap}}$ & Cu4-O6 &Cu1-O1    &Cu4-O6   &Cu1-O1&Cu1-O1&Cu1-O2\\   
&1.835(3)&2.010(20)&1.884 &1.957&1.836&1.854  \\

Cu$_{\mathrm{ch}}$-O$_{\mathrm{ch}}$($\parallel c$) &Cu4-O7 &-    &Cu4-O7   &-&Cu1-O4&-      \\
& {1.735(16)} &-&1.807 &-&1.876 &- \\

Cu$_{\mathrm{ch}}$-O$_{\mathrm{ch}}$($\parallel b$) & Cu4-O7 &-    &Cu4-O7   &-&Cu1-O4&-      \\
& {1.975(3)} &-&1.960 &-&1.955 &- \\

\hline

\end{tabular}
\label{TA3}
\end{table*}

\begin{figure*}[ht]
\begin{center}
\includegraphics[width=14cm]{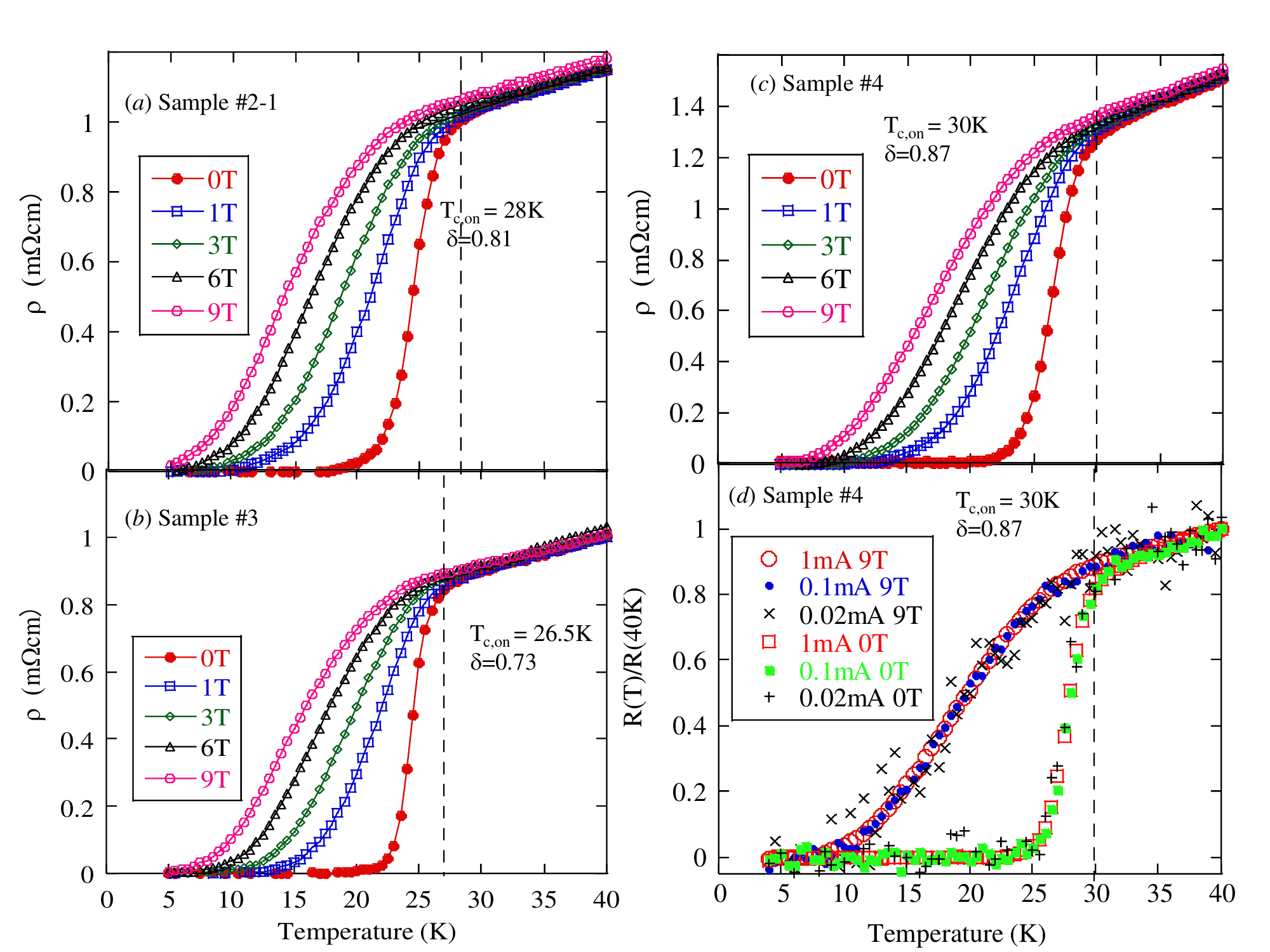}
\caption{Low temperature dependences of  electric resistivities  for the superconducting Pr$_{2}$Ba$_{4}$Cu$_{7}$O$_{15-\delta }$ measured at the several applied fields ($\mu_{0}H$=0, 1, 3, 6, and 9 T). ($a$) sample \#2-1 with  $\delta$=0.81, ($b$) sample \#3 with $\delta$=0.74, and  ($c$) sample \#4 with $\delta$=0.8. In  ($d$), the data for the  sample \#4  are taken at 0T and 9T for the applied current $I$=1, 0.1 and 0.02 mA .  Broken lines are guides for the eyes.  }
\label{RTH}
\end{center}
\end{figure*}

\begin{figure*}[ht]
\begin{center}
\includegraphics[width=14cm, pagebox=cropbox, clip]{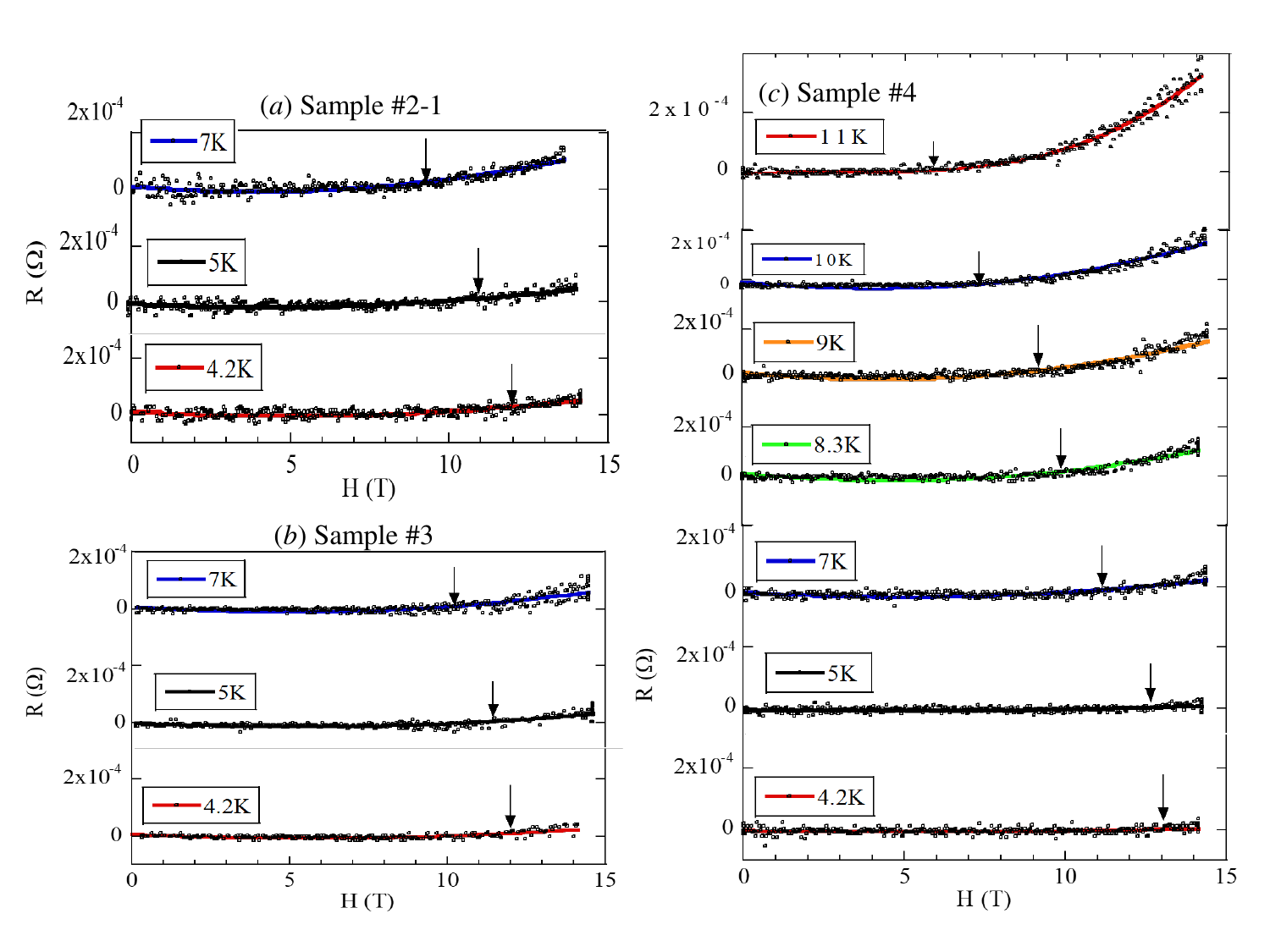}
\caption{Low-temperature magneto-resistance effect   as a function of magnetic field up to 14 T for the superconducting Pr$_{2}$Ba$_{4}$Cu$_{7}$O$_{15-\delta }$. ($a$) sample \#2-1 and ($b$) sample \#3  at $T$ = 4.2 K, 5 K, and 7 K.  ($c$) sample \#4  at $T$= 4.2 K$\sim$11 K . The arrow symbols denote the critical fields $H\mathrm{_{c}}^{*}$, where  a zero-resistance state is violated  at selected temperatures upon increasing applied field. Lines are guides for the eyes. }
\label{RH1}
\end{center}
\end{figure*}

\section{Results and Discussion}
Figure \ref{CR} shows the schematic  projection along (100) of  the crystal structure of Pr$_{2}$Ba$_{4}$Cu$_{7}$O$_{15-\delta}$.   The CuO sinlgle-chain and double-chain blocks are alternatively stacked along the c-axis.   The two-phase Rietveld refinement results for  the as-sintered and oxygen-reduction samples of    Pr$_{2}$Ba$_{4}$Cu$_{7}$O$_{15-\delta}  $  ($\delta$ =0, 0.73, and 0.81) are listed in Table \ref{TA1}. The typical powder synchrotron X-ray diffraction data  on the superconducting sample Pr$_{2}$Ba$_{4}$Cu$_{7}$O$_{15-\delta}  $  ($\delta$ = 0.81) at room temperature are given in Fig. \ref{RI} .  The calculated and residual curves are also plotted using  RIETAN-FP program  for  two-phase  Pr$_{2}$Ba$_{4}$Cu$_{7}$O$_{15-\delta} $ and BaCuO$_{2} $ .  The mass fractions of Pr247 and BaCuO$_{2}$  were estimated to be $ 96.0 \% : 4.0 \% $. 
 
  Here, we give some comments on the atomic position parameters of Pr$_{2.2}$Ba$_{3.8}$Cu$_{7}$O$_{15-\delta }$ ($\delta$=0.8) listed in Table \ref{TA2} \cite{BO88}.  In the present Rietveld refinement, 
 we assume that the Ba1 and Ba2 sites  are partially replaced by Pr ions with same occupancy rates of 0.05 in the 123 and 124 blocks, respectively.  For the Pr123 compound, it has been reported that  the Pr rich and Ba poor compositions Pr$_{1+x}$Ba$_{2-x}$Cu$_{3}$O$_{7-\delta} $ are stable against the stoichiometric one  from the single phase diagram studies \cite{SO03}.  These findings strongly suggest that  to form the Pr123 single phase,  the Ba sites must have partial Pr ions.   This  tendency  is  naturally  extended to the Pr124 and Pr247 systems.  The occupancy rates  at the oxygen sites O8 and O9 in the 123 block are set to be 0.1, to match  the oxygen defect.

  A  vacuum reduction treatment of the as-sintered samples introduces  oxygen defects, resulting in carrier doping effect  associated with the appearance of superconducting phases.  It is well known that higher $T_{c}$ samples are closely related to larger oxygen defects accompanied by the $c$-axis elongation of Pr247.  The variations of the lattice parameters due to oxygen removal are almost similar to the previous ones\cite{YA05}.  In Table \ref{TA3},  the selected bond lengths for Pr$_{2}$Ba$_{4}$Cu$_{7}$O$_{15-\delta }$ determined by the powder  synchrotron X-ray diffraction patterns are listed.
Comparing  the interatomic distances  for  the superconducting and non-superconducting samples, we detected no clear differences in all the bond lengths between both the samples. As for the Cu$_{\mathrm{ch}}$-O$_{\mathrm{ch}}$(Cu4-O7 $\parallel c$) bond length corresponding to the inter-chain distance along the c-axis  in the 124 block,  it seems to be  slightly shorter in the $\delta$ = 0.81 sample with  $T_{c,on}$=28 K  than that in the non-superconducting as-sintered sample.  The normal to superconducting phase diagram of a CuO double- chain has been investigated on the basis of Tomonaga-Luttinger liquid theory\cite{SA05}. 
In the $n-t_{pp}$ binary phase diagram,  $n$ is electron density in the $d$ band of  the CuO double chain and $t_{pp}$ is hopping energy between nearest neighbor oxygen sites occupied by  2$p_{\sigma}$ orbitals.
While the system initially remains in the normal phase, the electron carrier doping effect 
causes a phase transition of it  from the normal to superconducting state.
Furthermore,  a shrinkage of the distance between two chains of a CuO double chain gives rise to an increase in the hopping energy $t_{pp}$.    As a result of the heavy oxygen defects,    it is expected that the double-chain system in the normal phase is varied  diagonally towards the superconducting phase  on the $n-t_{pp}$ plane, realizing double-chain driven  superconductivity.

\begin{figure*}[ht]
\begin{center}
\includegraphics[width=14cm, pagebox=cropbox, clip]{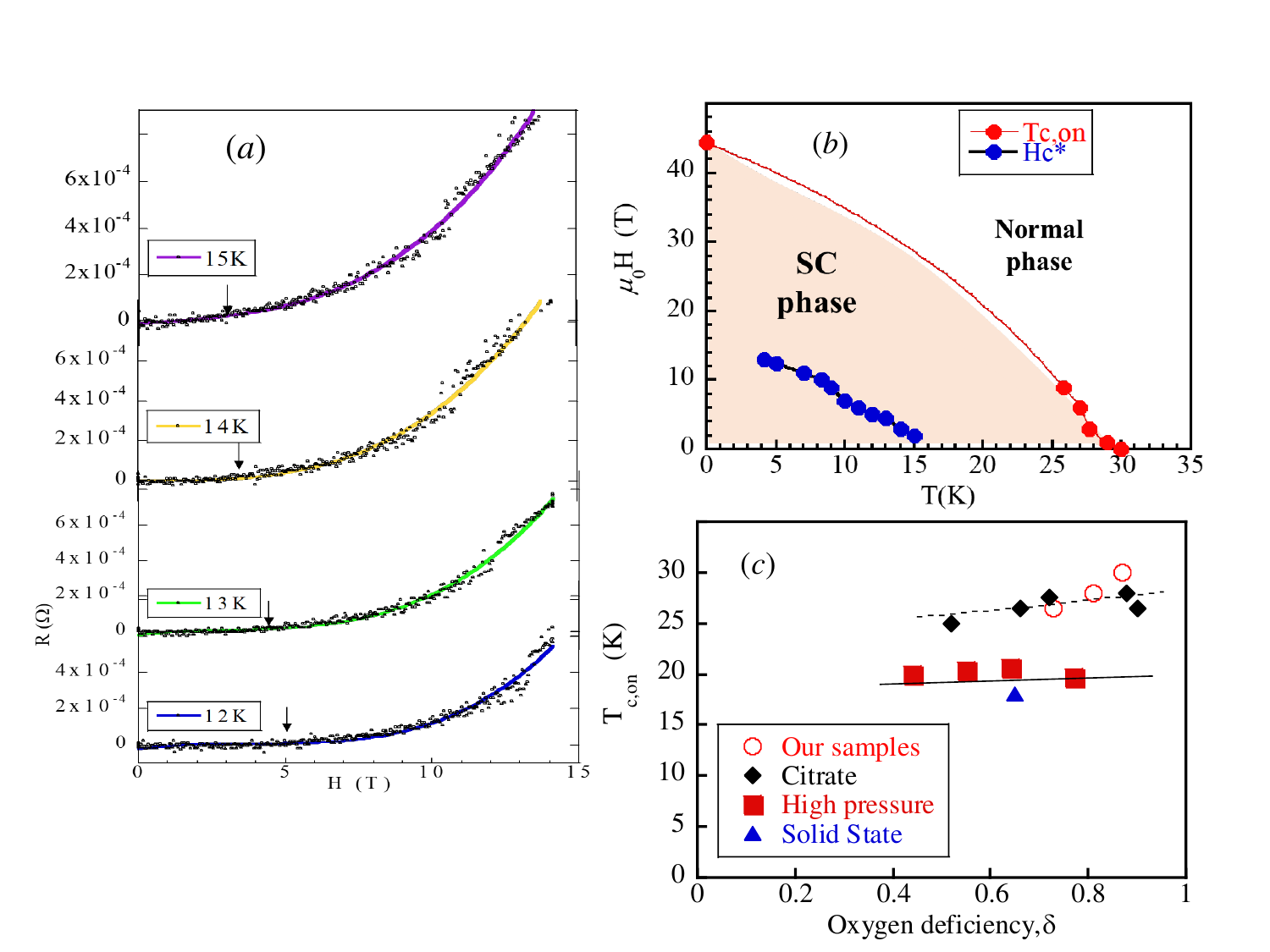}
\caption{(Color online) ($a$) Low-temperature magneto-resistance effect   as a function of magnetic field up to 14 T for the superconducting Pr$_{2}$Ba$_{4}$Cu$_{7}$O$_{15-\delta }$ (sample \#4)  at $T$=12$\sim$ 15 K. ($b$) Superconducting high-field phase diagram.   $T\mathrm{_{c,on}}$($H$) and $H\mathrm{_{c}}^{*}$($T$)   were taken from the temperature scan and magnetic field scan data.   
 ($c$) $T\mathrm{_{c, on}}$ vs oxygen deficiency for our samples ({$\bigcirc$}). For comparison, {\large $\blacklozenge$},  {\large $\blacksquare$}, and {\large $\blacktriangle$ } correspond to   the data of Hagiwara et al.\cite{HA08},  Ishikawa et al.\cite{IS09}, and Nishioka et al.\cite{NI22} ,respectively.  Lines are guides for the eyes. }
\label{PD}
\end{center}
\end{figure*}

Next,  in Fig. \ref{RT}, we show  the temperature dependences of  electric resistivities  for the superconducting Pr$_{2}$Ba$_{4}$Cu$_{7}$O$_{15-\delta }$.  The low-temperature magnetization curves measured under a zero field cooling (ZFC) scan  are also  plotted, to clarify a bulk superconductivity.  In addition,   we examined the magnetic field effect on the superconducting phase  of these three samples with different oxygen defects, as shown in Figures \ref{RTH}, \ref{RH1} and \ref{PD}($a$).  
First of all,  the reduction heat  treatment increased  the  onset temperature of superconductive transition $T\mathrm{_{c, on}}$  from 26.5 K at $\delta$ =0.7, through 28.0 K at $\delta$ =0.81 one, up to  30.0 K at $\delta$ =0.87.   From our magnetization data for the  three samples,   their superconducting volume fractions $f_\mathrm{SC}$ reached over 50 \%.
Furthermore, a zero-resistance state of these samples was observed below $T\mathrm{_{c, zero}}$=20$\sim$22 K at 0 T. 
On the other hand,  the room-temperature resistivity was enhanced from $\sim12$ m
$\Omega $ cm to  $\sim26$ m$\Omega $ with increasing the oxygen defect.  
This behavior is probably related to the highly  insulating state  
of  the CuO$_{2}$ planes due to the removal of  single-chain site oxygen in 123 block, as  suggested in the oxygen content study on the Pr123 compound \cite{LO90} .  
 In the PrBa$_{2}$Cu$_{3}$O$_{7-\delta }$ system,  it was reported that   the room-temperature resistivity is increased  by about three orders of magnitude from $\delta =$0 to $\delta \sim$0.5.  
 The present  Pr247 reduced samples 
showed the higher residual resistance ratios (RRR), $\rho$ (RT)/ $\rho$ (30K), ranging from 10 to 12,  as pointed out in our previous study\cite{HO21}.  
Assuming that a model of parallel resistors along the $b$ axis consists of CuO chain and CuO$_{2}$ plane resistances,  it is reasonable to consider the contribution of  a doped double chain alone to $\rho$(T), because of ignoring electric conducting path in a highly insulating plane. 
 
The application of external magnetic field up to 9 T caused widely broadening of the  resistive transition curves accompanied by disappearance of  a zero resistance state. 
$T\mathrm{_{c, on}}$ seems to be  stable with respect to the applied field, in spite of  a rapid degradation of  $T\mathrm{_{c, zero}}$. 

We examined the resistive transition profiles under 9 T  as a function of the electric current ranging from 0.02 mA to 1 mA.  In our present experiment (Fig. \ref{RTH} ($d$)) \cite{CH13},  there were no significant differences in the resistive drops among  0.02, 0.1 and 1mA.  
It is not made clear  that the motion of magnetic flux due to Lorenz force affects these broadening behaviors, because of  the applied  current densities   limited to lower values of  $\sim $ 10 to 40 mA/cm$^{2}$.
Our results are reminded  of  the  magnetic field effect on the  superconducting phases in the hole doped two-leg ladder compound Sr$_{2}$Ca$_{12}$Cu$_{24}$O$_{4}$ and high-$T\mathrm{_{c}}$ cuprate  (La$_{1-x}$Sr$_{x}$)$_{2}$ CuO$_{4}$\cite{NA05,KI89}. 

From the low-temperature  electric resistivity data for the superconducting sample \#4 under several applied fields ($\mu_{0}H$=0, 1, 3, 6, and 9 T) , we  obtained  the onset critical temperature as a function of  magnetic field $T\mathrm{_{c,on}}$($H$). 
Using the the Werthamer-Helfand-Hohenberg (WHH) formula in the dirty limit, $\mu_{0}H_\mathrm{c2}(0)=-0.69T_\mathrm{c}(dH_\mathrm{c2}(T)/dT)|_{T=T_{c}}$\cite{WE66},   the upper critical field at $T$= 0 K,    $H_\mathrm{c2}$(0)  was  evaluated to be  $\sim45$ T.  According to $H_\mathrm{c2}$=$\phi_\mathrm{0}/2\pi \xi^2$, we obtained the Ginsburg-Landau coherence length    $\xi \sim 27$ \AA,  which is comparable to the in-plane coherence length for  (La$_{1-x}$Sr$_{x}$)$_{2}$ CuO$_{4}$ single-crystal thin films\cite{SU91}. 
To further establish the superconducting high-field phase diagram, we determined from the low-temperature MR effect the resistive critical field  as a function of temperature  $H\mathrm{_{c}}^{*}$($T$) , where a zero-resistance state is violated  at selected temperatures upon increasing applied field. 
For the present samples,  $H\mathrm{_{c}}^{*}$  is estimated to be 12-13 T at 4.2 K. 
In the resistive crossover region from  $T\mathrm{_{c,on}}$($H$)  to   $H\mathrm{_{c}}^{*}$($T$)  line,  
it is hard to distinguish  between dissipation due to a melting of vortex lattice and an intrinsic effect associated with the  double-chain induced superconductivity.
The higher critical fields are probably  attributed to  the improved sample preparation process  using the three-zone electric furnace,  compared with the lower  $H\mathrm{_{c}}^{*}$  value ($\sim $4 T at 4 K) for the previous samples with higher $T\mathrm{_{c, on}} =$ 26.5 K \cite{KU16}.   Additionally, the quenched treatment just after vacuum reduction annealing contributes to the enhanced superconducting properties for the present samples. 
These procedures  are related to further improvements of weak links between the superconducting grains.

Finally, in Fig. \ref{PD} ($c$),  we summarize  $T\mathrm{_{c, on}}$ vs oxygen deficiency for our samples, compared with the data of several other groups \cite{HA08,IS09,NI22}.
It is noted that  the higher  $T\mathrm{_{c}}$ samples around $\sim 26$ K are synthesized by using the citrate pyrolysis based precursors under the different heat treatment conditions.   On the other hand,  the lower  $T\mathrm{_{c}}$ samples are annealed under ambient  and high pressure oxygen  atmospheres  by using a solid-state reaction method for mixed fine powders.  It is true that the value of   $T\mathrm{_{c}}$ is closely dependent on the oxygen defects.  However,  the differences  between two-types of the lower ad higher  $T\mathrm{_{c}}$ samples  are difficult to explain  on the basis of  the oxygen occupancy at the CuO single chain sites only.  
According to high-resolution electron microscopy studies \cite{HA08,CH13} on Pr247, there exit stacking faults in  the -123-124-123-124- (-S-D-S-D-) regular sequence along the $c$ axis as shown in Fig.\ref{CR}. 
Here,  S and D correspond to the single and double CuO chain blocks.
For instances,  the stacking faults such as the -123-123-123-124- (-S-S-S-D-)  irregular sequence  are reported for the citrate pyrolysis samples.   For the Y247 system\cite{KA98,GU95},  it is also  concluded that  the difference in  $T\mathrm{_{c}}$ values is not attributed to the oxygen occupancy but micro structures.
Through TEM measurements  on different samples with two-types of preparation methods,  stacking faults along the $c$-axis are observed more frequently in a polymerized-complex  sample with $T\mathrm{_{c}}$ = 93 K  than in a solid-state reaction sample with $T\mathrm{_{c}}$ = 65 K.  
Accordingly, we expect  that higher  $T\mathrm{_{c}}$ samples driven by CuO double chain superconductivity  are related to  stacking faults,
 although  CuO$_{2}$ based layered superconductivity of Y247  is quite different from that of  the present Pr247 system. 
 
 \section{Summary}
 We demonstrated the lattice structures and  the superconducting phases of metallic double-chain based cuprate   Pr$_{2}$Ba$_{4}$Cu$_{7}$O$_{15-\delta}$ exhibiting higher $T\mathrm{_{c}}$.  
  After the oxygen heat treatment on the citrate pyrolysis precursors in the thee-zone temperature controlled furnace, their reduction treatment followed by a rapidly cooling procedure caused higher $T\mathrm{_{c}}$ samples with 28$\sim$30 K.
We carried out  the powder synchrotron X-ray diffraction measurements  on the as-sintered and vacuum reduction samples at Spring-8 beamline.  In particular,   the crystal structural parameters for 
the superconducting sample   ($\delta$ = 0.81) were investigated  using  RIETAN-FP program  for  two-phase  Pr$_{2}$Ba$_{4}$Cu$_{7}$O$_{15-\delta} $ and BaCuO$_{2}$.  

We examined the magnetic field effect on the superconducting phase  of these three samples with different oxygen defects  ($\delta$ =0.73, 0.81 and 0.87) and established  the superconducting magnetic field-temperature  phase diagram from the low-temperature MR effect.  
For the $\delta$ = 0.87 sample  with $T\mathrm{_{c}}$ $\sim$30 K,    $H\mathrm{_{c}}^{*}$ was estimated to be 13 T at 4.2 K.    $H_\mathrm{c2}$(0)  was  evaluated to be  $\sim45$ T,   resulting in  the Ginsburg-Landau coherence length    $\xi \sim 27$ \AA.   
When  oxygen deficiency  dependence  on $T\mathrm{_{c, on}}$ for our samples   was  compared with the data of several other groups \cite{HA08,IS09,NI22},  better superconductive samples driven by CuO double chains  have a close relationship with  stacking faults along the $c$-axis.

The authors are grateful for M. Nakamura for his assistance in PPMS experiments at Center for Regional
Collaboration in Research and Education, Iwate University. M. M thanks Prof. M. Hagiwara for his valuable comments.
The synchrotron radiation experiments were performed at the BL02B2 of SPring-8.

\end{document}